\documentstyle[prl,aps]{revtex}
\begin{document}
\draft
\title{Breaking democracy with non renormalizable mass terms}
\author{Joaquim I. Silva-Marcos\cite{juca}}
\address{Theoretical Physics Division, CERN, \\
CH-1211 Geneva 23, Switzerland}
\maketitle

\begin{abstract}
The exact democratic structure for the quark mass matrix, resulting from the
action of the family symmetry group $A_{3L}\times A_{3R}$, is broken by the
vaccum expectation values of heavy singlet fields appearing in non
renormalizable dimension 6 operators. Within this specific context of
breaking of the family symmetry we formulate a very simple ansatz which
leads to correct quark masses and mixings.
\end{abstract}


{\bf Introduction}

One of the outstanding problems in particle physics is the problem of the
fermion masses and mixings. In the standard model (SM), which, most likely,
is an effective theory at low energy, these physical quantities are computed
from the Yukawa couplings. With regard to the quarks, one can have, in
principle, for the 3 families of the up and down sector, 18 complex Yukawa
couplings. This gives us a total of 36 parameters from which one has to
extract the 10 physical quantities: 6 quark masses, 3 mixing angles and a CP
violating complex phase.

To reduce this large amount of parameters, or even to find possible
relations between the quark masses and mixings \cite{rel}, one is lead to
seek, e.g., for new symmetries which act among the family structure \cite
{symrel}. Another approach is to postulate, ab initio, ans\"atze for the
Yukawa couplings which lead to phenomenological viable patterns \cite{pat} 
\cite{heav} \cite{dem} \cite{sym}. The hope is to find some hint about a
symmetry principle behind the mechanism of fermion mass generation. In the
literature, there are, grosso modo, two classes of ans\"atze. Those which
are formulated in a ''heavy'' weak basis \cite{pat} \cite{heav}, where one
of the Yukawa couplings of each sector is much larger then the other, and
ans\"atze which are formulated in the ''democratic'' weak basis \cite{dem} 
\cite{sym}, and where all Yukawa couplings of each sector are almost equal
to each other.

In this paper, we present a very simple but phenomenological correct pattern
within the democratic weak basis. In our approach, the exact democratic
structure is generated through the action of the family symmetry group $%
A_{3L}\times A_{3R}$, where $A_3\subset S_3$ is the subgroup of even
permutations. This group is then broken by the vacuum expectation values
(v.e.v.) of heavy singlet fields appearing only in non renormalizable
dimension 6 operators. The idea is, therefore, that the exact democratic
structure is broken by contributions from higher order operators arrizing in
the scenario (which will not be discussed here) of some unified theory at a
large scale $M=M_{GUT}-M_{Pl}$ \cite{oper}. Within this specific context of
breaking of the $A_{3L}\times A_{3R}$ family symmetry we formulate a very
simple ansatz which leads to correct quark masses and mixings.

{\bf General framework}

As known, the discrete family symmetry $A_{3L}\times A_{3R}$ generates (and
not necessarily $S_{3L}\times S_{3R}$ as one often finds) the democratic
mass matrix: 
\begin{equation}
\label{e1}\Delta =\left[ 
\begin{array}{ccc}
1 & 1 & 1 \\ 
1 & 1 & 1 \\ 
1 & 1 & 1 
\end{array}
\right] 
\end{equation}
Our model consists of the usual $SU(2)_L\times U(1)_Y$ SM Higgs doublet $%
\phi $, the left-handed quark doublets $L_i$ and right-handed singlets $R_i$
(which here represent either the right handed up quarks $u_{R_i}$ or the
right handed down quarks $d_{R_i}$). Both $L_i$ and $R_i$ transform
trivially with respect to the $A_3$ family symmetry, i.e., the family
indices transform as 
\begin{equation}
\label{e2}
\begin{array}{lll}
(1)=e\ ;~ & (123)=a\ ;~ & (132)=b 
\end{array}
\end{equation}
$A_3$ is isomorf to $Z_3$. This can be easily checked from its
multiplication table: $a^2=b$ and $a\ b=e$, which leads to $a^3=b^3=e$ (and
also $b^2=a$). Then, the lowest dimension mass term in the Lagrangean which
is invariant under this independent interchange of the left and right-handed
fields is

\begin{equation}
\label{e3}\lambda \ (\overline{L_1}+\overline{L_2}+\overline{L_3})\ \phi \
(R_1+R_2+R_3) 
\end{equation}
and one gets the democratic mass matrix.

In order to change the democratic structure, we introduce now two
independent Higgs ($A_3$ family) triplets $X_i$ and $Y_i$: one transforming
(in the same way and) together with the left-handed and the other with the
right-handed fields. Under $SU(2)_L\times U(1)_Y$ they are singlets. One can
form three independent $A_3$ invariant combinations: 
\begin{equation}
\label{e4}
\begin{array}{c}
Z_1=a_1\ b_1+a_2\ b_2+a_3\ b_3 \\ 
Z_2=a_1\ b_3+a_2\ b_1+a_3\ b_2 \\ 
Z_3=a_1\ b_2+a_2\ b_3+a_3\ b_1
\end{array}
\end{equation}
where the $(a_i,b_i)$ either stand for the independent $A_3$ partners $(%
\overline{L_i},X_i)$ or for the $(R_i,Y_i)$. The next to lowest dimension
(and non-renormalizable) mass terms, are, e.g., combinations like $(%
\overline{L_1}X_1+\overline{L_2}X_2+\overline{L_3}X_3)\ \phi \
(R_1Y_2+R_2Y_3+R_3Y_1)$. An extra $Z_2$ symmetry is needed to avoid the
combinations $\overline{L}X\phi R$ or $\overline{L}\phi YR$. Please notice
also that the exact democratic structure appearing in the Lagrangean, as a
result of the combination in Eq. (\ref{e3}) is in fact invariant under $%
A_{3L}\times A_{3u_R}\times A_{3d_R}$, because it is possible to transform
the right-handed up quark fields independently from the right-handed down
quark fields. However, with the introduction of the new singlets fields $X_i$
and $Y_i$, this larger symmetry is no longer valid as the $Y_i$ fields
require that the $u_{R_i}$ and $d_{R_i}$ transform simultaneously. Thus,
here, we have an exact $A_{3L}\times A_{3R}$ family symmetry.

The whole mass term in the Lagrangean, including the lowest and the relevant
next to lowest order dimension mass operator, will be 
\begin{equation}
\label{e4a}\lambda \ (\overline{L_1}+\overline{L_2}+\overline{L_3})\ \phi \
(R_1+R_2+R_3)+\lambda _{mk}\ \frac{Z_m^{(\overline{L},X)}}M\ \ \phi \ \frac{%
Z_k^{(R,Y)}}M
\end{equation}
where the $Z_k$ were defined in Eq.(\ref{e4a}), e.g., $Z_2^{(\overline{L}%
,X)}=(\overline{L_1}X_3+\overline{L_2}X_1+\overline{L_3}X_2)$, and where $M$
is the heavy mass where the large scale structure of the unified theory
becomes apparent. The $A_3$ symmetry of the singlet fields is broken when
they acquire the following v.e.v.'s \cite{derm}: 
\begin{equation}
\label{e4b}
\begin{array}{c}
\left( <X_1>,<X_2>,<X_3>\right) =(0,0,V_X) \\  
\\ 
\left( <Y_1>,<Y_2>,<Y_3>\right) =(0,0,V_Y)
\end{array}
\end{equation}
The quark mass matrix, thus obtained, for each sector, will then be of the
form: 
\begin{equation}
\label{e5}M^{\circ }=\lambda v\ \left( \left[ 
\begin{array}{ccc}
1 & 1 & 1 \\ 
1 & 1 & 1 \\ 
1 & 1 & 1
\end{array}
\right] \ +\ \left[ 
\begin{array}{ccc}
a_{11} & a_{12} & a_{13} \\ 
a_{21} & a_{22} & a_{23} \\ 
a_{31} & a_{32} & a_{33}
\end{array}
\right] \ \right) 
\end{equation}
where $a_{ij}=(\lambda _{ij}/\lambda )\ (V_XV_Y\ /\ M^2)$ and which is, as
one can clearly see, not democratic any more. In fact, all family symmetries
have been broken. The heavy singlets get their v.e.v.'s at a scale which, at
least, should be smaller than the mass scale $M$. Thus the $a_{ij}$ are
smaller than $1$. Because of the large scale $M$, the other dimension 6
operators involving only quark field combinations should be even (much)
smaller, as the v.e.v's from the heavy singlets do not contribute to these
terms.

{\bf The ansatz}

Within this type of democracy breaking context, we shall consider the
specific case where, compared to the three parameters $%
(a_{13},a_{31},a_{32}) $, all other $a_{ij}$ are small. This is a natural
limit, in the sense that we are not demanding any special relations between
the $a_{ij}$ like, e.g., in the ansatz of Fritzsch \cite{fri} or the cases
classified by Ramond Roberts and Ross \cite{pat} where $M_{12}^{\circ
}=M_{21}^{\circ }$ and $M_{23}^{\circ }=M_{32}^{\circ }$. Taking the limit
where the small $a_{ij}\rightarrow 0$ we obtain the following
(dimensionless) asymmetric mass matrix, 
\begin{equation}
\label{3a}M=\left[ 
\begin{array}{ccc}
1 & 1 & 1+a_{13} \\ 
1 & 1 & 1 \\ 
1+a_{31} & 1+a_{32} & 1 
\end{array}
\right] 
\end{equation}
Parametrizing $a_{31}$, $a_{32}$ and $a_{13}$ as follows,

\begin{equation}
\label{3aa}
\begin{array}{l}
a_{31}=q\ e^{i\ \alpha }+r\ e^{i\ \beta } \\ 
a_{32}=q\ e^{i\ \alpha } \\ 
a_{13}=r\ e^{i\ \beta }\ (1+\varepsilon \ e^{i\ \gamma })
\end{array}
\end{equation}
does not add anything to our ansatz, as $a_{31}$, $a_{32}$ and $a_{13}$
remain independent. However, it is very useful to study the phenomenological
implications of Eq. (\ref{3a}). To do this, we shall first concentrate on a
simplification of Eq. (\ref{3a}). As an example, we take the case where $%
\varepsilon =0$ and $\alpha ,\beta =\pi /2$. One gets, 
\begin{equation}
\label{3b}M[
\begin{array}{c}
\varepsilon =0, \\ 
\alpha ,\beta =\pi /2
\end{array}
]\approx \left[ 
\begin{array}{lll}
1 & 1 & e^{i\ r} \\ 
1 & 1 & 1 \\ 
e^{i\ (q+r)} & e^{i\ q} & 1
\end{array}
\right] =\left[ 
\begin{array}{lll}
1 & 1 & e^{i\ (q+r)} \\ 
1 & 1 & e^{i\ q} \\ 
e^{i\ (q+r)} & e^{i\ q} & e^{i\ q}
\end{array}
\right] \cdot K_R
\end{equation}
where we have used the approximation $1+i\ x\approx e^{ix}$. The unitary
matrix $K_R=$ diag $(1,1,e^{-i\ q})$ is non-relevant and can be absorbed in
a transformation of the right-handed quark fields. The mass matrix on the
right-hand side of Eq. (\ref{3b}) is exactly one of the familiar symmetric
cases described in the USY hypothesis of Ref. \cite{sym} with two
dimensionless parameters. Obviously, the diagonalization matrix elements,
such as $U_{12}$ and $U_{23}$, depend on these. In a first order
approximation, it was found that $U_{12}=(\sqrt{3}/2)\ (r/q)$ and $U_{23}=(2%
\sqrt{2}/9)\ q$. Since $q$ and $r$ depend on the mass ratios through the
(approximate) relations, $q=(9/2)(m_2/m_3)$ and $r=3(3m_1m_2)^{1/2}/m_3$,
the phenomenological formulas $U_{12}=(m_1/m_2)^{1/2}$and $U_{23}=\sqrt{2}%
(m_2/m_3)$ are obtained \cite{dem} \cite{sym}. Notice the precise (and
peculiar) cancellation of the numerical factors.

Let us now present an analysis of the general mass matrix in Eq. (\ref{3a}).
We shall assume that $\varepsilon =o(m_2/m_3)\ll 1$. This is rather a
special choice in parameter space, i.e., it is not natural (in the sense
explained above), because in that case $a_{31}-a_{32}\approx a_{13}$ ; it is
a choice motivated by predictability. We shall not go into the details of
solving the characteristic equations, which involve the mass ratios of the
quarks of the physical relevant square mass matrix; that was done in Ref. 
\cite{sym}. Defining $H=\ M\ M^{\dagger }/t$, where $t=tr(H)$ is such that $%
tr(H)\equiv 1$, one obtains eigenvalues that respect exact, $\lambda
_1+\lambda _2+\lambda _3=1$, and approximate relations: 
\begin{equation}
\label{3c}
\begin{array}{lll}
\lambda _1=\frac{m_1^2}{m_3^2}~;\quad  & \lambda _2=\frac{m_2^2}{m_3^2}%
~;\quad  & \lambda _3=1
\end{array}
\end{equation}
From the characteristic equations one finds, in first order, approximate
values for $q=(9/2)(m_2/m_3)$ and $r=3(3m_1m_2)^{1/2}/m_3$. Then, using an
iteration method starting with these initial approximate values, one finds
expressions for $q$ and $r$ as series in mass ratios.

\begin{equation}
\label{3d}
\begin{array}{l}
r= 
\frac{3\sqrt{3m_1m_2}}{m_3}\cdot \left[ 1+\frac 32\left( \frac{m_2}{m_3}%
\right) \cos (\alpha )-\frac 12\varepsilon \cos (\gamma )+\ldots \right] \\  
\\ 
q=\frac 92\ \frac{m_2}{m_3}\cdot \left[ 1-\sqrt{\frac{4m_1}{3m_2}}\cos
(\alpha -\beta )+\ldots \right] 
\end{array}
\end{equation}
The phases $\alpha $ , $\beta $ and the $\varepsilon $ are free parameters;
they are not determined by the mass ratios. We shall come to this later.

After introducing these relations into the square mass matrix $H$, one
computes the eigenvectors, also as a series in the mass ratios. The
diagonalization matrix $U$ is calculated in the heavy weak basis. In this
weak basis all matrix elements of are small except $H_{33}$, and only the
relevant contributions of $H_u$ and $H_d$ to $V_{CKM}$ are present. Thus the
irrelevant parts, which cancel out in the Cabibbo-Kobayashi-Maskawa matrix
(product), 
\begin{equation}
\label{e6}V_{CKM}=U_u^{\dagger }\cdot U_d 
\end{equation}
are absent. In this way, both $U_u$ and $U_d$ are both near $\openone$. The
heavy weak basis is defined in the following way,

\begin{equation}
\label{e7}
\begin{array}{l}
H_u\longrightarrow H_u^{
\text{Heavy}}=F^{\dagger }\cdot H_u\cdot F \\ H_d\longrightarrow H_d^{\text{%
Heavy}}=F^{\dagger }\cdot H_d\cdot F 
\end{array}
\qquad ;\qquad F=\left[ 
\begin{array}{ccc}
\frac 1{\sqrt{2}} & \frac 1{\sqrt{6}} & \frac 1{\sqrt{3}} \\ \frac{-1}{\sqrt{%
2}} & \frac 1{\sqrt{6}} & \frac 1{\sqrt{3}} \\ 0 & \frac{-2}{\sqrt{6}} & 
\frac 1{\sqrt{3}} 
\end{array}
\right] 
\end{equation}
One finds for the diagonalization matrix elements $U_{12}$ and $U_{13}$,

\begin{equation}
\label{e8}
\begin{array}{l}
|U_{12}|= 
\sqrt{\frac{m_1}{m_2}}\left[ 1-\frac{m_1}{2m_2}+\frac{m_2}{m_3}\cos \alpha +%
\frac \varepsilon 2\cos \gamma +\ldots \right] \\  \\ 
|U_{13}|=\frac 1{\sqrt{2}}\frac{\sqrt{m_1m_2}}{m_3}\left[ 1-\frac{m_2}{2m_3}%
\cos \alpha +\frac \varepsilon 2\cos \gamma +\ldots \right] \  
\end{array}
\end{equation}
where the next to leading order terms are of small influence. For the
elements $U_{23}$ and $U_{31}$ one obtains next to leading order terms which
are somewhat larger, 
\begin{equation}
\label{e8a}
\begin{array}{l}
|U_{23}|= 
\sqrt{2}\frac{m_2}{m_3}\left[ 1-\sqrt{\frac{3m_1}{4m_2}}\cos (\alpha -\beta
)+\ldots \right] \\  \\ 
|U_{31}|=\frac 3{\sqrt{2}}\frac{\sqrt{m_1m_2}}{m_3}\left[ 1-\sqrt{\frac{m_1}{%
3m_2}}\cos (\alpha -\beta )+\ldots \right] 
\end{array}
\end{equation}
Approximate relations hold 
\begin{equation}
\label{e9}
\begin{array}{ll}
|U_{13}|=\frac 12\ |U_{23}\ U_{12}|\quad ;\qquad & |U_{31}|=3\ |U_{13}| 
\end{array}
\end{equation}

{\bf CP violation and a numerical example}

In this section we describe the CP violation and the masses and mixings of a
numerical example of the ansatz in Eq. (\ref{3a}). We find that CP violation
is mainly restricted by the range which, within our framework, is possible
to have for the up quark mass $m_u$.

It is clear that, on the one hand, for general mass matrices $M_{u,d}$ of
type Eq. (\ref{3a}), the CP violation depends, crucially, on the complex
phases $\alpha _{u,d}$ and $\beta _{u,d}$, which are free parameters,
independent of the mass ratios, and which for a specific numerical (ansatz)
example still have to be fixed. Obviously, for $\alpha ,\beta =k\pi $, there
is no CP violation. On the other hand, if $M_u$ and $M_d$ are real, we find
for the $V_{CKM}$ matrix element 
\begin{equation}
\label{e10}|V_{us}|=\left| \sqrt{\frac{m_d}{m_s}}\pm \sqrt{\frac{m_u}{m_c}}%
\right| 
\end{equation}
where the $\pm $ sign depends on the relative signs of $r$ and $q$ (i.e., if 
$\alpha ,\beta =k\pi $) for the up and down sector. Combining the
experimental limits on $m_d/m_s$, $m_s$ and $m_c$, one can only accommodate
the experimental value of $|V_{us}|=0.2196(23)$ \cite{pdg} in Eq. (\ref{e10}%
) if one takes a very small value for $m_u\leq 1\ MeV$ or even $m_u=0$.
However, when $\alpha ,\beta \neq k\pi $, the $\pm $ sign in Eq. (\ref{e10})
is replaced by a complex phase factor such that 
\begin{equation}
\label{e11}|V_{us}|=\left| \sqrt{\frac{m_d}{m_s}}+e^{i\delta }\ \sqrt{\frac{%
m_u}{m_c}}\right| 
\end{equation}
and it is possible to accommodate a larger value for $m_u$ \cite{delta}.
Clearly for our ansatz, CP violation is closely related to this problem,
i.e., it depends also on the $\alpha $'s and $\beta $'s and subsequently on $%
\delta $ which is a function of these. Numerically, we have found that CP
violation, given by $|J_{CP}|=|Im(V_{us}V_{cb}V_{cs}^{\star }V_{ub}^{\star
})|$, is large when also $\delta $ mod $\pi $ is large. Thus, a larger value
for $m_u$ can only be accommodated if one takes values for $\alpha ,\beta
\neq 0$ mod $\pi $ such that $\delta $ mod $\pi $ is large and this results
in a large value for the CP violation parameter (and vice versa). In order
to find (ansatz) examples with sufficient large CP violation, it is useful
to have an expression for $\delta $.

Let us compute $\delta $ in a first order approximation. Writing the
eigenvalue equation of each quark sector as $H=U\cdot D\cdot U^{\dagger }$,
where $H$ is given in the heavy basis of Eq. (\ref{e7}) and $D=$ diag$%
(\lambda _1,\lambda _2,\lambda _3)$ contains the eigenvalues of $H$, one
obtains (using the unitarity of $U$), the exact relations 
\begin{equation}
\label{e12}
\begin{array}{l}
(\lambda _2-\lambda _1)\ U_{12}U_{32}^{\star }+(\lambda _3-\lambda _1)\
U_{13}U_{33}^{\star }=H_{13} \\ 
(\lambda _2-\lambda _1)\ U_{22}U_{32}^{\star }+(\lambda _3-\lambda _1)\
U_{23}U_{33}^{\star }=H_{23} \\ 
(\lambda _2-\lambda _1)\ U_{12}U_{22}^{\star }+(\lambda _3-\lambda _1)\
U_{13}U_{23}^{\star }=H_{12}
\end{array}
\end{equation}
Using Eqs. (\ref{3c}, \ref{e8}, \ref{e8a}) and choosing $U_{33}$ real (this
is always possible), we find from the first two equations that the complex
phases of $U_{13}$ and $U_{23}$ are approximately equal to those of $H_{13}$
respectively $H_{23}$. Computing $H$ in the heavy basis with the
parametrization of Eq. (\ref{3aa}), one finds (for $\alpha $ and $\beta $
not too close to $k\pi $) in a first order approximation 
\begin{equation}
\label{e13}
\begin{array}{ll}
H_{13}=\frac 1{3\sqrt{6}}\ \ r\ e^{i\beta }~;\quad  & H_{23}=
\frac{2\sqrt{2}}9\ \ q\ e^{i\alpha }~\quad \Longrightarrow  \\  &  \\ 
U_{13}=|U_{13}|\ e^{i\beta }~~;\quad  & U_{23}=|U_{23}|\ e^{i\alpha }
\end{array}
\end{equation}
In addition, from Eqs. (\ref{3c}, \ref{e8}, \ref{e8a}) one obtains $\lambda
_2/\lambda _3=|U_{13}U_{23}^{*}|/|U_{12}U_{22}^{\star }|$. Thus 
\begin{equation}
\label{e14}|\lambda _2U_{12}U_{22}^{*}|=|\lambda _3U_{13}U_{23}^{\star }|=%
\frac{m_1m_2}{m_3^2}\ \sqrt{\frac{m_2}{m_1}}
\end{equation}
holds and because $H_{12}$ $=-r^2/18\sqrt{3}=-\sqrt{3}m_1m_2/2m_3^2$ is
smaller than this (in absolute value), we may conclude from the third
relation in Eq. (\ref{e12}) that, aside from a factor $\pi $%
\begin{equation}
\label{e15}\arg (U_{12}U_{22}^{*})=\arg (U_{13}U_{23}^{\star })
\end{equation}
Unitarity also tells us that, in this approximation, $\arg
(U_{11}U_{21}^{*})=\arg (U_{12}U_{22}^{\star })$. Finally, putting together
all these phase relations for the up and down sector, we get (aside from any
factors $\pi )$ 
\begin{equation}
\label{e16}\delta =(\alpha _d-\beta _d)-(\alpha _u-\beta _u)
\end{equation}
With this expression, we can now choose suitable combinations for $\alpha
_{u,d}$ and $\beta _{u,d}$ to have a large $\delta $ (mod $\pi $) in order
to account for suitable large values for CP and $m_u$.

Next we give a numerical example,where we take $\varepsilon _{u,d}=0$ and
the simplest combinations for $\alpha _{u,d}$ and $\beta _{u,d}$ to obtain a
large $\delta $. The mass matrices of both sectors are (as explained) of
type 
\begin{equation}
\label{e17}M=c\ \left[ 
\begin{array}{lll}
1 & 1 & 1+r\ e^{i\beta } \\ 
1 & 1 & 1 \\ 
1+q\ e^{i\alpha }+r\ e^{i\beta } & 1+q\ e^{i\alpha } & 1
\end{array}
\right] 
\end{equation}
Example with $\delta =-\pi /3$, where $\alpha _d=\alpha _u=\beta _u=0$ and
only $\beta _d=\pi /3$ (extra $\pi $ factors are put in as signs in the $q$%
's and $r$'s) and 
\begin{equation}
\label{e18}
\begin{array}{ll}
r_d=-3.259\times 10^{-2} & r_u=9.368\times 10^{-3} \\ 
q_d=0.1254 & q_u=1.463\times 10^{-2} \\ 
c_d=2\ GeV & c_u=133\ GeV
\end{array}
\end{equation}
which at $1$ $GeV$ correspond to,%
$$
\begin{array}{ll}
m_d=7.39\ MeV & m_u=3.73\ MeV \\ 
m_s=186\ MeV & m_c=1.38\ GeV \\ 
m_b=6.2\ GeV & m_t=400\ GeV
\end{array}
$$
give 
\begin{equation}
\label{e19}|V_{CKM}|=\left[ 
\begin{array}{lll}
0.9748 & 0.2229 & 0.0037 \\ 
0.2225 & 0.9740 & 0.0414 \\ 
0.0124 & 0.0397 & 0.9991
\end{array}
\right] \quad ;\qquad \frac{|V_{ub}|}{|V_{cs}|}=0.0896
\end{equation}
and $|J_{CP}|=1.8\times 10^{-5}$. To obtain a large value for $|J_{CP}|$ one
would expect that a value for $\delta =\pm \pi /2$ would be more suitable.
However, $J_{CP}$ depends also on other order contributions which are of
significant importance. Numerically, we have found that $\delta =\pm \pi /3$
gives the largest values for $|J_{CP}|$.

{\bf Concluding remarks}

We have shown that the exact democratic structure for the quark mass
matrices, resulting from the action of the family symmetry group $%
A_{3L}\times A_{3R}$, can be totally broken by the effects of non
renormalizable dimension 6 operators adding a small perturbation to this
structure. Within this context, we formulate a unique ansatz: one of the
simplest deviations from democracy, requiring a minimum of parameters, and
which predicts the well known phenomenological mixings in terms of quark
mass ratios. We have also shown that CP violation is determined by a simple
combination of complex phases of these parameters. A numerical
ansatz-example is given in good agreement with experiment.

\begin{acknowledgements}
I would like thank E.K. Akhmedov for the careful reading of this paper.
I would also like thank the hospitality of CERN Theory Division, 
where part of this work was done.
This work was partially supported by Funda\c c\~ao
para a Ci\^encia e Tecnologia of the Portuguese Ministry of Science and
Technology.
\end{acknowledgements}


\begin{references}
\bibitem[*]{juca}  E-mail address: Joaquim.Silva-Marcos@cern.ch

\bibitem{rel}  J.L.Rosner, hep-ph/9610222; T. Kobayashi and Z.-Z. Xing, Mod.
Phys. Lett. A 12 (1997) 561; Y. Achiman, hep-ph/9812389; B. Stech, Phys.
Lett. B 465 (1999) 219; H. Fritzsch, Nucl. Phys. B 556 (1999) 49; K.
Harayama, Prog. Theor. Phys. 103 (1999) 135; A. Mondragon and E.
Rodriguez-Jauregui, Phys. Rev. D 61 (2000) 113002; H. Fritzsch and Z.-Z.
Xing, Prog. Part. Nucl. Phys. 45 (2000) ; P.H. Frampton, P.Q. Hung and M.
Sher, Phys. Rept. 330 (2000) 263

\bibitem{symrel}  C.D. Froggatt and H.B. Nielsen, Nucl. Phys. B 147 (1979)
277; L. Ib\'a{\~n}ez and G. G. Ross, Phys. Lett. B 332 (1994) 100; Y.P. Yao,
hp-ph/9507207; E.K. Akhmedov, M. Lindner, E. Schnapka and J.W.F. Valle,
Phys. Rev. D 53 (1996) 2752; T. Ito, N. Okamura and M. Tanimoto, Phys. Rev.
D 58 (1998) 077301; C.H. Albright and S.M. Barr, Phys. Lett. B 452 (1999)
287; G. Altarelli and F. Feruglio, Phys. Lett. B 451 (1999) 388; I.S.
Sogami, H. Tanaka and T. Shinohara, Prog. Theor. Phys. 101 (1999) 707; K.
Hagiwara and N. Okamura, Nucl. Phys. B 548 (1999) 60; M. Fukugita, M.
Tanimoto and T. Yanagida, hep-ph/9903484; S.L. Adler, Phys. Rev. D 59 (1999)
015012; L. Lavoura, Phys. Rev. D 61 (2000) 077303; R. Kitano and Y. Mimura,
hep-ph/0008269; Q. Shafi and Z. Tavartkiladze, Phys. Lett. B 482 (2000) 145;
G. Altarelli, F. Feruglio and I. Masina, hep-ph/0007254

\bibitem{pat}  P. Ramond, R.G. Roberts and G.G. Ross, Nucl. Phys. B 406
(1993) 19

\bibitem{heav}  G. Branco and J. I. Silva-Marcos, Phys. Lett. B, 331 (1994)
390; T. Ito, Prog. Theor. Phys. 96 (1996) 1055; E. Takasugi and M.
Yoshimura, Prog. Theor. Phys. 98 (1997) 1313; E. Takasugi, Prog. Theor.
Phys. 98 (1997) 177; A. Rasin, Phys. Rev. D 58 (1998) 096012; M.
Baillargeon, F. Boudjema C. Hamzaoui and J. Lindig, hep-ph/9809207; D.
Falcone, Mod. Phys. Lett. A 14 (1999) 1989; T.K. Kuo, S.W. Mansour and G.H.
Wu, Phys. Rev. D 60 (1999) 093004; D. Falcone and F. Tramontano, Phys. Rev.
D 61 (2000) 113013

\bibitem{dem}  G.C. Branco, J.I. Silva-Marcos and M.N. Rebelo, Phys. Lett. B
237 (1990) 446; G.C. Branco and J.I. Silva-Marcos, Phys. Lett. B 359 166
(1995) 174; G.C. Branco, D. Emmanuel-Costa and J.I. Silva-Marcos, Phys. Rev.
D 56 (1997) 107; T. Teshima and T. Sakai, Prog. Theor. Phys. 97 (1997) 653;
P.M. Fishbane and P.Q. Hung, Mod. Phys. Lett. A 12 (1997) 1737; {\it ibid }%
Phys. Rev. D 57 (1998) 2743; P.M. Fishbane and P. Kaus, Z. Phys. C 75 (1997)
1; I.S. Sogami, K. Nishida, H. Tanaka and T. Shinohara, Prog. Theor. Phys.
99 (1998) 281; T. Teshima and T. Asai, hep-ph/0009181

\bibitem{oper}  Non renormalizable operators were studied within the context
of different symmetry groups in, e. g., S. F. King, Phys. Lett. B 325 (1994)
129; B.C. Allanach and S.F. King, Nucl. Phys. B 456 (1995) 57; {\it ibid }B
459 (1996) 75; B.C. Allanach, S.F. King, G.K. Leontaris and S. Lola, Phys.
Lett. B 407 (1997) 275; See also the review on the SM, G. Altarelli,
hep-ph/9912291.

\bibitem{derm}  E. Derman and H.S. Tsao, Phys. Rev. D 20 (1979) 1207

\bibitem{fri}  H. Fritzsch, Phys. Lett. B 70 (1977) 436

\bibitem{sym}  J.I. Silva-Marcos, Phys. Lett. B 443 (1998) 276

\bibitem{pdg}  Particle Data Group

\bibitem{delta}  H. Fritzsch, hep-ph/9501316;{\it \ ibid, } hep-ph/9909303;
H. Fritzsch and Z.-Z. Xing, Phys. Lett. B 413 (1997) 396
\end{references}
\end{document}